\documentstyle[twocolumn,prl,aps]{revtex}
\begin{document} 
\title{
Shakeup Effects on Photoluminescence from the Wigner Crystal
}
\author{D.Z. Liu}
\address{
Center for Superconductivity Research,
Department of Physics, University of Maryland,
College Park, MD 20742
}
\author{H.A. Fertig}
\address{
Department of Physics and Astronomy, University of Kentucky,
Lexington, Kentucky 40506-0055
}
\author{S. Das Sarma}
\address{
Department of Physics, University of Maryland,
College Park, MD 20742
}
\address{\rm (Submitted to Physical Review Letters on 13 June 1994)}
\address{\mbox{ }}
\address{\parbox{16cm}{\rm \mbox{ }\mbox{ }
We develop a method to compute shakeup effects on photoluminescence
in a Wigner crystal from localized holes.  Our method treats the
lattice electrons and the tunneling electron on an equal footing, and
uses a quantum-mechanical calculation of the collective modes
that is realistic throughout the Brillouin zone.  We find that
shakeup produces a series of sidebands that may be identified
with maxima in the collective mode density of states,
and definitively distinguishes the crystal state from a liquid state.
We also find a shift in
the main luminescence peak, that is associated with lattice
relaxation in the vicinity of a vacancy.}}
\address{\mbox{ }}
\address{\parbox{16cm}{\rm PACS numbers: 78.20.Ls, 72.20.Jv, 73.20.Dx}}
\maketitle

\makeatletter
\global\@specialpagefalse
\def\@oddhead{REV\TeX{} 3.0\hfill Das Sarma Group Preprint, 1994}
\let\@evenhead\@oddhead
\makeatother

Sixty years ago, Wigner\cite{wigner} pointed out that an electron
gas will undergo a zero-temperature, quantum phase transition into
a crystalline phase as the density is lowered.
Forty-five years later, the first convincing evidence of an electron
crystal was presented for a system of electrons on a He surface\cite{grimes}.
The electron densities attainable in this fashion are extremely low,
however, making this an unattractive system for observing the quantum
phase transition.  Semiconductors are much more
attractive systems in this sense, because one has great control
over the electron densities, through dopant concentrations.
A particularly good candidate for observing the Wigner crystal (WC)
is the two-dimensional electron gas (2DEG),  as realized in modulation
doped semiconductors.  Samples of this type are now available
with such high quality that the electron groundstate is not
necessarily dominated by disorder.  The possibility
of observing the WC is further enhanced by the application of a strong
perpendicular magnetic field, which quenches the kinetic energy,
and allows the formation of a crystal state at higher densities
(for which disorder effects are less important) than
would be possible without it.

Experimental evidence for the WC in 2DEG's has accumulated over the
last several years\cite{rev}. One probe that has produced much
intriguing data is photoluminescence (PL), in which either
a valence band hole\cite{clark} or a hole bound to an
acceptor\cite{buhmann,kukushkin} recombines with an electron in
the 2DEG, producing a characteristic photon spectrum.
A mean-field analysis\cite{pl} of the latter type of experiment showed
that the PL spectrum has, in principle, characteristic
signatures of the WC:  a
``Hofstadter butterfly''\cite{hofstadter}
spectrum for the case of weak interactions between the
electrons and the hole, and a characteristic shift in the
PL spectrum upon melting of the crystal.

In this Letter, we go beyond the mean-field approximation,
to examine shakeup effects on the PL spectrum; i.e.,
we will examine how the collective mode spectrum of the
WC (which, at long wavelengths, corresponds to the
classical phonon spectrum) ,
and the fact that some of these modes may be excited
in the electron-hole recombination process, modify the results
of the mean-field theory.  We will consider in detail only the case of
a localized hole\cite{buhmann,kukushkin}.
Our method is purely quantum-mechanical, and treats both the
tunneling electron and the other lattice electrons on
the same footing. Furthermore, we employ a quantum
treatment of the collective modes to realistically account for
contributions both from small and large wavevector collective
excitations of the lattice.  Since we are working in the
strong magnetic field limit, we consider only excitations
within the lowest Landau level (LLL).
Our main results are: (1) Shakeup effects
shift the main PL peak to higher energies
than found in a mean-field treatment\cite{pl}.
(2) The Hofstadter spectrum is eliminated from the PL spectrum,
even in the case of weak electron-hole interactions\cite{kukushkin} (although
we will argue below that it survives in the itinerant
hole case\cite{pl}).  The sudden shift of the PL spectrum upon
melting, by contrast, survives even when shakeup is included.
(3) Phonon sidebands appear
that correspond to maxima in the phonon density of states (DOS);
some (but not all) of these sidebands are results of van Hove
singularities in the DOS, and so are characteristic of an
ordered WC state.  For the case of
weak electron-hole interactions\cite{kukushkin},
we {\it do not} see these sidebands
in the liquid state, so that phonon satellites uniquely
distinguish between a liquid and a solid state.  Interestingly,
for stronger electron-hole interactions, a shakeup satellite
persists even above the melting temperature.  We expect this
sideband to lose oscillator strength relative to the main
peak, either with increasing
temperature or decreasing electron-hole interaction strength.
The latter may be accomplished by examining PL from several
samples with different acceptor - 2DEG setback distances \cite{kukushkin}.

Examples of our calculated PL spectra are shown in Fig. 1 for filling fractions
$\nu=1/5$ and $\nu=2/7$, for electron density
$N_s=6\times10^{10}cm^{-2}$.  Our hole is assumed to be strongly localized,
and located $250\AA$ from the electron plane.
For the case of no electron-hole interaction [Fig. 1(a)],
at low temperature, a well-defined shakeup peak may be seen approximately
2 meV below the main PL peak; a second very weak satellite
is observed approximately 3.5meV below the main peak.
The origins of these peaks may be understood in terms of
the phonon DOS, which is illustrated in Fig. 2.
A van Hove singularity, arising from zone-edge phonons, appears as
a strong double peak near 0.4meV.  Two other peaks
may be seen near 1.2 meV and 1.9 meV.
There are weak sidebands associated with each of
these peaks in the PL spectrum.  The precise interpretation
of these peaks is unclear; however, it has been speculated that
these represent vacancy-interstitial excitations\cite{rc}.
We point out that it is {\it crucial} to use a fully quantum
mechanical treatment of the collective excitations of the
lattice to observe these higher order satellites;
classical treatments of the
phonons\cite{johansson,bonsall}
do not produce these unusual excitations.

\begin{figure}
 \vbox to 6.0cm {\vss\hbox to 8cm
 {\hss\
   {\includegraphics{fig1.ps}
   }
  \hss}
 }
\caption{
PL spectra for (a)$\nu=1/5$, $T=0$, no electron-hole
interaction; (b)$\nu=1/5$, $T=0$, with electron-hole
interaction;(c)$\nu=2/7$, $T=0$, no electron-hole
interaction.  Inset: PL spectra for (a) and (b), with $T$ just
above the melting temperature. In order to distinguish different
spectra, they were seperated by $250$ units.
}
\end{figure}

It should also be noted that the splitting between the main
PL peak and the first sideband is actually larger than the energy
at which the van Hove singularity in the phonon DOS appears.
The reason for this is that there is a strong self-energy
renormalization due to the phonons in the main PL peak.
Physically, this arises because the final state
of the crystal contains a vacancy, which is lowered in energy
by a distortion of the lattice -- i.e., by allowing the electrons
surrounding the vacancy to relax inward.  The self-energy shift
accounts for this lowering in energy of the final state of the WC, and
leads to an upward shift of the main PL peak.
Our calculation shows that the final states in
which phonons are excited are
not nearly so strongly renormalized by lattice relaxation
effects, leading to the increased splitting between
the main PL peak and the sidebands.


The inset to Fig. 1 illustrates the PL spectrum in the melted state
both without and with an electron-hole interaction.
In the former case, there is no phonon sideband present.  This is
necessarily so, because in the melted phase, the density is uniform,
and there are no collective modes in the LLL \cite{rc}.
By contrast, when an
electron-hole interaction is present, there is a non-uniform electron density
near the hole, allowing some local collective modes to persist even
above the melting temperature.  With further increase in temperature,
or increased setback between the hole and the 2DEG,
the oscillator strength of this mode will significantly decrease.

Fig. 1(c) illustrates our results for $\nu=2/7$ in the absence of
electron-hole interactions.  As can be seen,
except for a change in energy scale caused by changing the
magnetic field,
the lineshape is essentially
identical to the case of $\nu=1/5$.  This contrasts sharply with
the results found in the mean-field approximation \cite{pl}, where
\begin{figure}
 \vbox to 6.0cm {\vss\hbox to 8cm
 {\hss\
   {\includegraphics{fig2.ps}
   }
  \hss}
 }
\caption{
Collective mode DOS for $\nu=1/5$, $T=0$.
}
\end{figure}
\noindent
without electron-hole interactions, a filling
$\nu=p/q$ generally yields $p$ distinct lines for a localized hole.
While the splittings are so small in that situation that they
are difficult in practice to resolve, evidently shakeup effects
wipe out this structure even in principle.  We note that for the
case of an itinerant hole, we expect these characteristic
splittings to survive shakeup effects.
The reason is that (neglecting excitonic effects), the PL spectrum
is related to the product of Green's functions for the
hole and the electron.  For localized holes, only the latter
has poles in the form of a Hofstadter spectrum at the
mean-field level, which are wiped out by shakeup effects.
However, for itinerant holes, the hole Green's function also
has poles of the Hofstadter form, which are unaffected by shakeup.
Thus, one should in principle be able to
\onecolumn
\noindent
identify this characteristic spectrum of bands and gaps that is unique to
a WC in a magnetic field in itinerant hole experiments\cite{c1}.

We now present an outline of how these results were derived;
a more detailed description will be presented in a future
publication\cite{fut}.  Previously\cite{pl}, it was shown that
if we consider a superlattice of holes, with $n_e$ electrons
and one hole per unit cell,
the PL intensity may quite generally be written in the form
$
P(\omega) \propto {\rm Im} [R(\omega+i\delta )],
$
where
$R(\omega) =
\bigl(n_h \Omega /2\pi l_0^2\bigr)\sum_{\bf G} R({\bf G},\omega)
e^{-G^2l_0^2/4},$
and we have approximated the core-hole wavefunctions as delta-functions,
$n_h$ is the density of holes, $\Omega$ the volume of the system, and
the vectors ${\bf G}$ are the reciprocal lattice vectors of the superlattice.
The magnetic length $\l_0=\Bigl(\hbar c/ eB\Bigr)^{1/2}$ will
be set to unity in the remainder of this paper.
$R({\bf G},\omega)$ is defined as
$R({\bf G},\omega) = \bigl(1/g\bigr)\sum_{X}e^{-iG_xX+iG_xG_y/2}
R_{ii}(X,X-G_y;\omega),$
where $g$ is the Landau level degeneracy,
$X$ is the guiding center quantum number, and
$$
R_{ij}(X_1,X_2;i\omega_n)
=-\int_0^{\beta}<T_{\tau}a_{X_1}(\tau)c_i(\tau)
c_j^{\dag}a_{X_2}^{\dag}>e^{i\omega_n \tau} d\tau,
$$
with $a_X^{\dag}$ creating an electron in state $X$ and $c_i^{\dag}$
creating a hole in the unit cell $i$, and $\beta$ is
the inverse temperature.  Working in the lowest Landau
level, the equation of motion for $R_{ij}({\bf G},\omega)$
may be written as
\begin{eqnarray}
{{\partial} \over {\partial \tau}} R_{ij}({\bf G},\tau)
&&=
<\rho({\bf G},\tau)> \delta_{ij} \delta (\tau)
-\epsilon_0  R_{ij}({\bf G},\tau)
-n_h \sum_{{\bf G}^{\prime}}
V({\bf G}^{\prime})e^{i {\bf G}^{\prime} \times {\bf G}/2 - G^{\prime 2}/4}
R_{ij}({\bf G}-{\bf G}^{\prime},\tau)\nonumber\\
&&
-{1 \over {\Omega}} \sum_{{\bf q} \neq 0} v({\bf q})
C_{ij}(-{\bf q},{\bf q}+{\bf G}) e^{i {\bf G} \times {\bf q} /2 - q^2/2}
-{1 \over {\Omega}} \sum_{{\bf q}}V({\bf q})
C_{ij}(-{\bf q},{\bf G}) e^{-i {\bf q} \cdot {\bf R}_i - q^2/4}.
\label{eqmot}
\end{eqnarray}
In Eq.(\ref{eqmot}),  $v({\bf q})$ and $V({\bf q})$ are the Fourier
transforms of the electron-electron and
electron-hole interactions, respectively, $\epsilon_0$ is the
energy of the localized hole, and the sum over ${\bf G}^{\prime}$
is only over reciprocal lattice vectors, while the sums over
${\bf q}$ are over all wavevectors.
${\bf R}_i$ specifies the position of the hole in the $i~th$ unit cell,
and $<\rho({\bf G},\tau)>e^{-G^2/4}$ is the expectation value of
a Fourier component of the electron density.  While this quantity
is independent of $\tau$ in the groundstate, it will be convenient
for later purposes to formally leave it as an argument of the density.
The method for computing these Fourier components has been
described elsewhere\cite{pl,rc}.
The correlation function $C_{ij}({\bf p}_1,{\bf p}_2)$
in the last two terms is just Fourier transformation of the following
correlation function:
\begin{equation}
C_{ij}(X_1X_2;X_3X_4;\tau^{\prime},\tau)
\equiv -g
<T_{\tau} a_{X_1}^{\dag}(\tau^{\prime})a_{X_2}(\tau^{\prime})
a_{X_3}(\tau)c_i(\tau)c_j^{\dag}(0)a_{X_4}^{\dag}(0)>.
\label{cij}
\end{equation}
Eq.(\ref{eqmot}) represents the first in an infinite series of equations
relating an $n$ particle Green's function to the $n+1$ particle
Green's function\cite{baym}.  In the mean-field approximation,
it is simplified by employing a Hartree-Fock (HF)
decomposition of Eq.(\ref{cij}), which converts
Eq.(\ref{eqmot}) into a self-consistent equation for $R_{ij}$~\cite{pl}.
To include shakeup effects, we instead extend this hierarchy to
one more level, writing down a self-consistent form
for $C_{ij}$ which explicitly contains the collective
mode excitations.  To carry out this program, it is convenient to
implicitly define a self-energy by substituting the last two terms in
Eq.(\ref{eqmot}) by
$-\sum_{{\bf G}^{\prime}}\int_0^{\beta}d\tau^{\prime}
\Sigma({\bf G},{\bf G}^{\prime};\tau-\tau^{\prime})
R_{ij}({\bf G}^{\prime},\tau^{\prime})$.
The HF approximation for the PL is equivalent
to taking $\Sigma({\bf G},{\bf G}^{\prime};\tau-\tau^{\prime})$
by $\Sigma^{HF}({\bf G},{\bf
G}^{\prime})\delta(\tau-\tau^{\prime})$, where
\begin{equation}
\Sigma^{HF}({\bf G},{\bf G}^{\prime})
 =
W({\bf G}-{\bf G}^{\prime})<\rho({\bf G}-{\bf G}^{\prime};\tau)>
e^{i {\bf G} \times {\bf G}^{\prime}/2}
+{1 \over {2\pi}}\sum_{{\bf q}}
V({\bf q})<\rho(-{\bf q},\tau)>
e^{-q^{2}/4-i{\bf q}\cdot{\bf R}_i} \delta_{{\bf G},{\bf G}^{\prime}},
\label{hfse}
\end{equation}
where $W$ is the sum of the direct and exchange Coulomb
potentials\cite{pl,rc}.
To generate a self-consistent equation for $C_{ij}$, we take a
functional derivative\cite{baym} of Eq.(\ref{eqmot}) with respect to a
spatially
periodic external potential $U({\bf p},\tau^{\prime})$.
In doing
this, it must be noted that a term directly coupling the
Green's function $R_{ij}$ to the external potential $U$ must
be added to Eq.(\ref{eqmot}), and the sums over reciprocal lattice vectors
must be extended to all wavevectors, because
$U$ is not in general commensurate with the lattice.
After an arduous calculation, the result expresses
$C_{ij}$ in terms of the density-density correlation function
$\chi({\bf G}_1+{\bf q},{\bf G}_2+{\bf q};\tau)
\equiv \chi_{{\bf G}_1 {\bf G}_2} ({\bf q};\tau)
=-g<T\tilde{\rho}({\bf G}_1+{\bf q},\tau)\tilde{\rho}(-{\bf G}_2-{\bf q};0)>$.
The Fourier transform of this response function contains poles
at the collective mode frequencies of the system; it is thus
through $\chi$ that shakeup effects are introduced.
We have computed this function numerically using a generalized
random phase approximation\cite{rc} which does not assume
small displacements of the lattice electrons
(as is necessary in classical approaches\cite{johansson}), and thus gives a
realistic dispersion relation for the collective modes across
the entire Brillouin zone. Upon substituting
the resulting $C_{ij}$ into Eq.(\ref{eqmot}),
one obtains the self-energy $\Sigma=
\Sigma^{HF}\delta(\tau-\tau^{\prime})+\delta \Sigma$, with
\newpage
\begin{eqnarray}
\delta \Sigma ({\bf G}_1,{\bf G}_2, \tau_1-\tau_2)~
&&= {1 \over {4 \pi}} \sum_{{\bf G}_1^{\prime}{\bf G}_2^{\prime}}
\int_{BZ} d^2{\bf q}
e^{i{\bf q} \times ({\bf G}_1-{\bf G}_2)/2
-i {\bf G}_1 \times {\bf G}_1^{\prime}/2
+i {\bf G}_2 \times {\bf G}_2^{\prime}/2}
W({\bf G}_1^{\prime}+{\bf q}) W({\bf G}_2^{\prime}+{\bf q})
\nonumber \\
&&
\times F({\bf G}_1-{\bf G}_1^{\prime}-{\bf q},
{\bf G}_2-{\bf G}_2^{\prime}-{\bf q}; \tau_1-\tau_2)
\chi_{{\bf G}_1^{\prime} {\bf G}_2^{\prime}} ({\bf q};\tau_1-\tau_2)
\nonumber \\
&&
+{1 \over {(2 \pi)^3}} \sum_{{\bf G}_1^{\prime}{\bf G}_2^{\prime}}
\int_{BZ} d^2{\bf q} e^{-({\bf G}_1^{\prime}+{\bf q})^2/4-
({\bf G}_2^{\prime}+{\bf q})^2/4}
V({\bf G}_1^{\prime}+{\bf q}) V({\bf G}_2^{\prime}+{\bf q})
\nonumber \\
&&
\quad \times F({\bf G}_1,{\bf G}_2;\tau_1-\tau_2)
\chi_{{\bf G}_1^{\prime} {\bf G}_2^{\prime}} ({\bf q};\tau_1-\tau_2)
\label{phse}
\end{eqnarray}
In Eq.(\ref{phse}), $\int_{BZ} d^2{\bf q}$
represents an integral over
wavevectors in the first Brillouin zone of the superlattice,
and $F$ is a generalized Green's function
satisfying the equation of motion
\begin{eqnarray}
{{\partial} \over {\partial \tau}}
F({\bf G},{\bf G}_1,\tau-\tau_1)
&&=
\delta_{{\bf G}{\bf G}_1}\delta (\tau-\tau_1)
-n_h \sum_{{\bf G}^{\prime}}
V({\bf G}^{\prime})e^{i {\bf G}^{\prime} \times {\bf G}/2 - G^{\prime 2}/4}
F({\bf G}-{\bf G}^{\prime},{\bf G}_1,\tau-\tau_1) \nonumber\\
&&-\epsilon_0  F({\bf G},{\bf G}_1,\tau-\tau_1)
-\sum_{{\bf G}^{\prime}}
\Sigma^{HF}({\bf G},{\bf G}^{\prime})
F({\bf G}^{\prime},{\bf G}_1,\tau-\tau_1),
\label{emf}
\end{eqnarray}
Eq.(\ref{emf}) may be solved using methods discussed previously\cite{pl,rc}.
With this expression, we are now able to
compute the PL intensity.  We substitute
$\Sigma=\Sigma^{HF}\delta(\tau-\tau^{\prime})+\delta\Sigma$
into Eq.(\ref{eqmot}), and Fourier transform this with respect to
imaginary time.
This means that a Fourier transform of Eq.(\ref{phse}) will
be necessary, leading to frequency summations of the form
(suppressing wavevector arguments)
$\sum_{i\omega_n} F(i\omega_n)\chi({\bf q};i\omega-i\omega_n)$.
To accomplish this, we represent $\chi$ as a sum over its
collective mode poles\cite{rc};
the frequency sums may be then computed using standard
methods\cite{baym}.
The computation of $\delta \Sigma$ is clearly the bottleneck in
this computation, since it requires two reciprocal lattice sums
and an approximate sum over the continuous wavevector ${\bf q}$.
We have accomplished this using
$469$ ${\bf q}$ points in the first Brillouin zone, for one and three
electrons per unit cell, which give very similar results.
Finally, once we have computed $\delta\Sigma$, it is straightforward
to substitute this into the frequency version of Eq.(\ref{eqmot}),
obtain $R({\bf G},\omega)$, and from there compute the PL
spectrum.

In summary, we have developed a method by which shakeup effects
in the PL spectrum of a WC from localized
holes may be computed, that treats the tunneling
electron and the lattice electrons on an equal footing,
and uses a fully quantum treatment of the collective
modes that is realistic over the entire Brillouin zone.
Our method is quite general, and should be applicable to other shakeup
problems where quantum fluctuations are important.
We find that the Hofstadter
spectrum found in a mean-field analysis of this
experiment is lost (although we expect it to survive
in itinerant hole experiments),
and is replaced by a series of sidebands due to
creation of phonons and other collective excitations of the
WC.  These sidebands are a unique signature of the WC, and
can in principle be used to distinguish between a liquid
and crystal state of the electrons.
We find that there is a sudden shift in the PL spectrum
upon melting of
the crystal.

\noindent {\it Acknowledgments.} The authors thank Dr. Ren\'e
C\^ot\'e for helpful discussions.  This work was supported by
the NSF, through Grant Nos. DMR 92-02255 and
DMR 91-23577, and by the US-ONR.
HAF acknowledges the support of the Alfred P. Sloan Foundation
and the Research Corporation, through a Cottrell Scholar Award.


\begin{references}
\bibitem{c1} This conclusion may be modified if the hole is close
enough to the electron layer that excitonic effects are important.
\end{references}
\end{document}